# A New Perspective to Node Influence Evaluation in Complex Network Using Subgraph Tr-Centrality

Auwal Tijjani Amshi,
School of Software, Nanchang Hangkong University, Nanchang, China.
auwalamshi@gmail.com

**Abstract: There is great significance in evaluating a node's Influence ranking in complex networks. Over the years, many researchers have presented different measures for quantifying node interconnectedness within networks. Therefore, this paper introduces a centrality measure called Tr-centrality which focuses on using the node triangle structure and the node neighborhood information to define the strength of a node, which is defined as the summation of Gruebler's Equation of the node's one-hop triangle neighborhood to the number of all the edges in the subgraph. Furthermore, we socially consider it as the local trust of a node. To verify the validity of Tr-centrality [1], we apply it to four real-world networks with different densities and shapes, and Tr-centrality has proven to yield better results.**

## I. INTRODUCTION

Many centrality measures have been introduced by different Researchers to evaluate node power and its importance in complex networks. Other researchers proposed different concepts of identifying importance, for instance, degree centrality [2] is based on the idea that important nodes must be the most active and have large numbers of direct links with other nodes, betweenness centrality [3] is a node can be simply quantified by the sum of probabilities that a node is on the path between two others, closeness centrality [4] is based on the idea that the sum of distances between important nodes and other nodes are the minimum, eigenvector centrality [5], and page-rank [6] and they both ranks a node as important if it is pointed to by other important nodes. Some researchers preferably consider Information centrality [7], which is based on the idea that the importance of a node is related to the ability of the network to respond to the deactivation of the node from the network. All the mentioned methods consider nodes important from either a global or local perspective.

In this paper, we construct a new centrality measure which is named "Tr-centrality" which computes the local strength of a node on its neighbors. Based on observation of the geometrical information a planar mechanism is somewhat similar to that of a planar network graph and by viewing the network graph as a planar mechanism, we can improve Gruebler's [8] equation to construct Tr-centrality and we consider the Tr-centrality of a node triangle neighborhood as that node quantify ranking.

We evaluate the performance of our proposed model by using four real-world network Datasets. Other six methods including triangle centrality [9], degree centrality, betweenness centrality, closeness centrality, eigenvector centrality, and Page Rank are also applied to the same selected networks for comparison. The results extensively prove the effectiveness of the proposed model.

## II. GRUEBLER'S EQUATION

The Gruebler's equation to calculate the degree of freedom of a planar mechanism reflects the fact that the structure of a planer mechanism and a planer graph are almost identical except the planer network graph doesn't have mobility properties and the Gruebler's equation is used to calculate the mobility and strength of a planar mechanism which is also called it degree of freedom.

Traditionally an n-link planar mechanism has a $3(N - 1)$ degree of freedom before any joints are connected. When constraints for all joints are subtracted from the total freedoms of the unconnected links, then the result is the mobility of the connection mechanism. [8] When we use $j_1$ to denote the number of single-degree-of-freedom pairs and $j_2$ for the number of two-degree-of-freedom pairs, the resulting mobility of a planar n-link mechanism is given by.

$$DOF = 3 * (N - 1) - 2j - j_2 \qquad (1)$$

The intuition behind the connection between Gruebler's equation and centrality is that Gruebler's equation is used to calculate the mobility (which can also be the strength) of a planar mechanism geometrical information [10] and is based on an observation of the geometrical information used when calculating the mobility of a planar mechanism is somewhat present in a planar network graph.

In graph theory, a planar graph is a graph that can be embedded in the plane, i.e., it can be drawn on the plane in such a way that its edges intersect only at their endpoints [11].

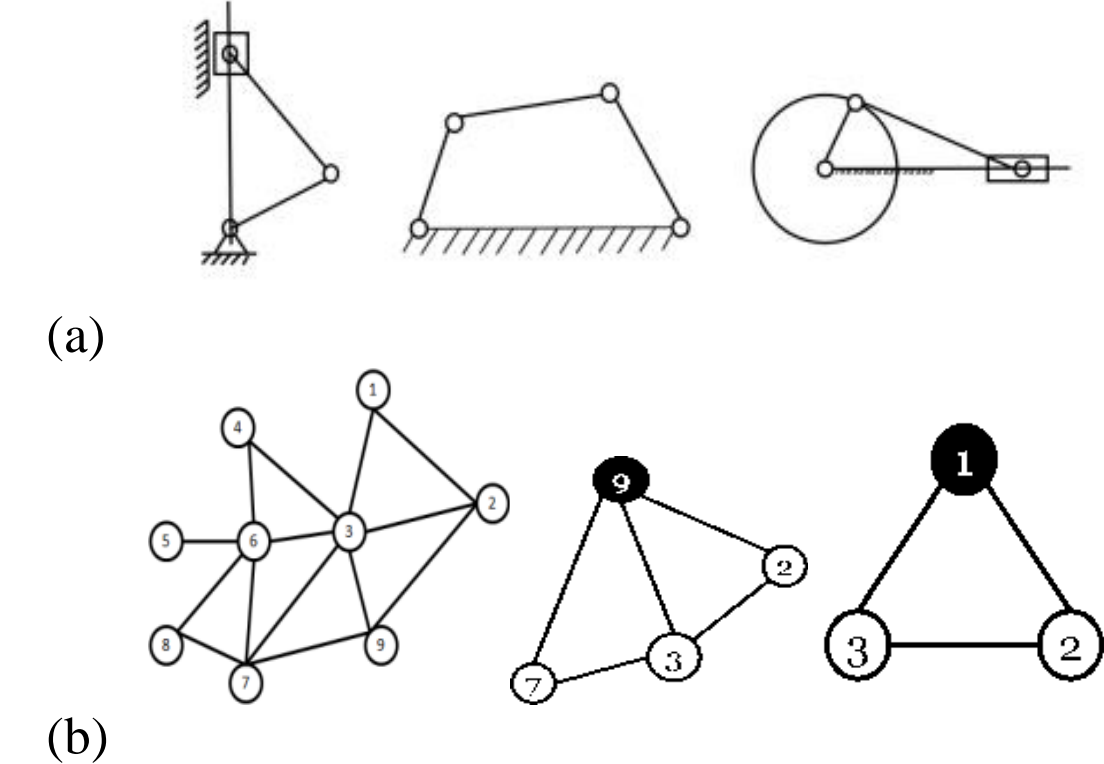

**Figure 1.** (a) illustrates an example of a Skeleton Outline planar mechanism. (b) Illustrates an example of a planar graph.

TABLE I

TWO BASIC SIMILARITY AND DIFFERENCES BETWEEN PLANAR MECHANISM AND PLANAR SUBGRAPH

| | Planar Mechanism | Planar Sub-graph |
|---|---|---|
| 1 | Join, Link | Node, Edges |
| 2 | Links are Mobile | Edges are not Mobile |

## III. TR-CENTRALITY

Given $G\ (V, E)$ undirected, unweighted graph with V nodes and E edges. We construct a sub-network represented by a subgraph $G_i$.

**Definition:** Given $\mathrm{G_i}$ let $\Gamma_i$ be the list of one-hop neighbors of node i and let $\gamma_i$ represent the list of one-hop triangle-connected neighbors of node i. For j and k neighbors of node i, if there exists a link between node j and k then append node j and k to list $\gamma_i$.

The subgraph degree centrality (sdeg) of node i, which can be denoted as $\mathrm{sdeg}_6$, is defined as:

$$\mathrm{sdeg}_{\ i} = |\gamma_i| \qquad (2)$$

Where $\gamma_i$ represents the list of one-hop triangle connected neighbors of node $i$.

**Algorithm I:** Subgraph Degree Centrality of Node i

```
Given (G, i)
let γi = []
    let Γi  = [onehop neigbors of node i ]
      for j and k ∈ Γi
        if ∃ ejk then
          b = j, k
          γi.append(b)
        return γi
```

To introduce Tr-centrality, we take Gruebler's equation and extend it in the following ways to evaluate node social strength of nodes in a complex network; we intentionally ignore using the global degree centrality (DC) of a node, because in this paper we are focusing on the local power of node. So we decompose the graph into triangular structured subgraph $\mathrm{G_i}$, based on definition 1, we obtain the local degree centrality of node $i$, which can be denoted as $\mathrm{sdeg_i}$ and by default $\mathrm{sdeg_i}$ is;

$$\mathrm{sdeg_i}\ \leq \mathrm{DC_i} \qquad (3)$$

Assuming $\mathrm{N_i}$ is the number of nodes in the node $i's$ one-hop triangle neighborhood, we can simply represent $\mathrm{N_i} - 1$ as $\mathrm{sdeg_i}$, as a consequence of the first section of equation (1), which is $(\mathrm{n}-1)$ and can also be replaced by $\mathrm{sdeg_i}$ and we replace the two unique constrain $j_1$ and $j_2$ as $N_i$ and $\mathrm{NT_i}$ (which is the number of triangles in node $i's$ neighborhood) respectively. The modified equations can be defined as [1]:

$$\mathrm{TC_i} = 3 \times (\mathrm{sdeg_i}) - (2 \times (\mathrm{N_i}) + \mathrm{NT_i}) + \sum_{i=1}^{M+1} \mathrm{sdeg_j} \qquad (4)$$

Where $N_i$, is the number of nodes in $i$ subgraph, while $\mathrm{NT}_{\ i}$ as the number of triangles in $i's$ subgraph and $\mathrm{sdeg}_{\ i}$ is the degree of $i$ in the subgraph $\mathrm{G_i}$. Meanwhile, M represents the one-hop triangle connected neighbors of node $i$.

The main deviation from Gruebler's degree of freedom equation is that our proposed equation will be used in graph theory to find the social power of nodes in a complex network. Therefore if a node has the highest Tr-centrality value in a network will be considered the most powerful or in other words most trusted.

To explain our equation we construct an example. Give $G(V, E)$, V = set of edges, and E = set of edges.

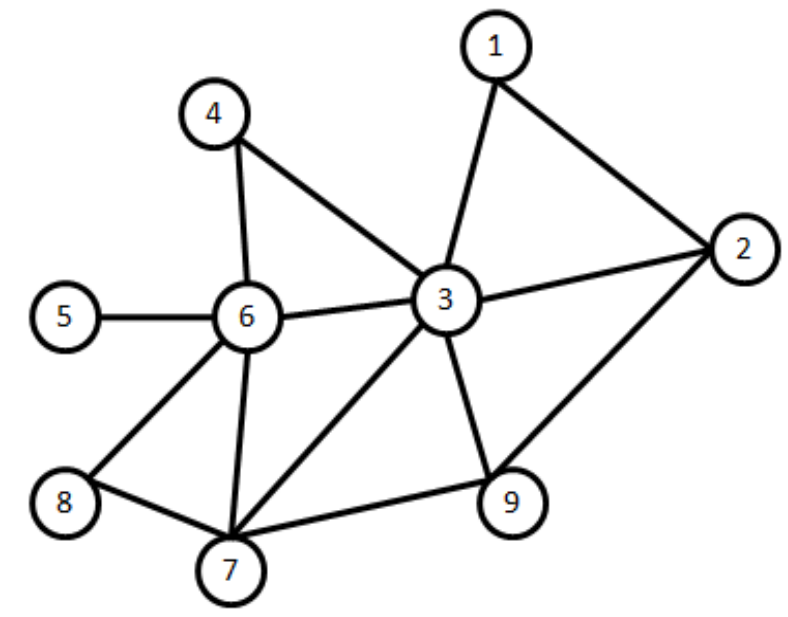

**Figure 2.** Experimental Graph

TABLE II
EXPERIMENTAL GRAPH INFORMATION

| Nodes | Degree Centrality (DC) | **sdeg** | Number of Triangle |
|---|---|---|---|
| Node 1 | 2 | 2 | 1 |
| Node 2 | 3 | 3 | 2 |
| Node 3 | 6 | 6 | 5 |
| Node 4 | 2 | 2 | 1 |
| Node 5 | 1 | 1 | 0 |
| Node 6 | 5 | 4 | 3 |
| Node 7 | 4 | 4 | 3 |
| Node 8 | 2 | 2 | 1 |
| Node 9 | 3 | 3 | 2 |

From Table II, we can see the difference in the value of the degree centrality $(DC_6)$ of node 6 and the value of the subgraph degree centrality $(\mathrm{sdeg}_6)$ of node 6 because node 5 is not a triangular connected neighbor of node 6, so it will not be part of subgraph $G_6$.

The first step is to construct the subgraph $G_i$ from graph G. Let's take node 1 and node 9 from the experimental graph as a sample.

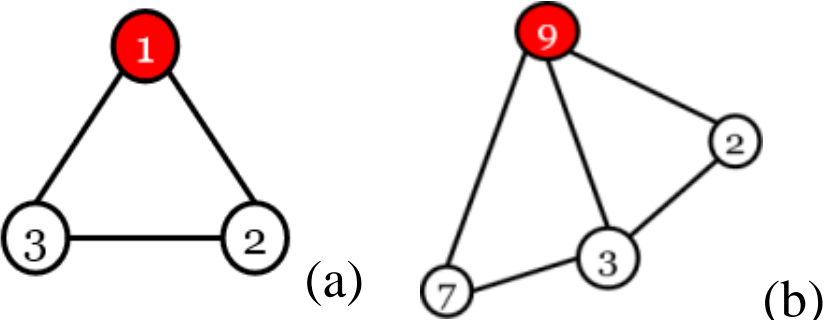

**Figure 3.** (a) Represent the subgraph $G_1$constructed by node 1 and (b) Represent subgraph $G_2$ constructed by node 9.

From equation (5)

$$\mathrm{TC_1} = 3 \times (\mathrm{sdeg_i}) - (2 \times (\mathrm{N_i}) + \mathrm{NT_i}) + \sum_{i=1}^{m+1} \mathrm{sdeg_j}\ = 5 * 0.01 = 0.05$$

$$\mathrm{TC_9} = 3 \times (\mathrm{sdeg_i}) - (2 \times (\mathrm{N_i}) + \mathrm{NT_i}) + \sum_{i=1}^{m+1} \mathrm{sdeg_j} = 9 * 0.01 = 0.09$$

We multiply the result by 0.01 to avoid returning a large number when dealing with large networks. We can observe that the proposed equation can distinguish differences in the social power of nodes.

TABLE III
RANKING RESULT OF EXPERIMENTAL GRAPH

| Nodes | 1 | 2 | 3 | 4 | 5 | 6 | 7 | 8 | 9 |
|---|---|---|---|---|---|---|---|---|---|
| TC Ranking | 6 | 4 | 1 | 7 | 9 | 2 | 3 | 8 | 5 |

## IV. Experiment, Results, and Discussion

To evaluate the performance of the proposed equation four real-world datasets [12] are used to verify its effectiveness.

TABLE IV
EXPERIMENTAL DATA

| Networks | Nodes | Edge |
|---|---|---|
| Zachary's karate club | 34 | 78 |
| Dolphins | 62 | 159 |
| Blogs Network | 112 | 425 |
| USAir97 | 332 | 2126 |

*Experiment 1: Node Ranking*

Comparison of top 5 nodes between proposed method TC with six other centrality measures, TR- number of Triangles, BC - Betweenness Centrality, CNC - Closeness Centrality, EC - Eigenvector Centrality, PR - Page Ranking, TC- Tr-Centrality.

TABLE V
TOP 5 NODES RANKING KARATE CLUB

| *Rank* | *TR* | *BC* | *CNC* | *EC* | *PR* | *TC* |
|---|---|---|---|---|---|---|
| 1 | 1 | 1 | 1 | 34 | 34 | 1 |
| 2 | 34 | 34 | 3 | 1 | 1 | 34 |
| 3 | 33 | 33 | 34 | 3 | 33 | 33 |
| 4 | 2 | 3 | 32 | 33 | 2 | 2 |
| 5 | 3 | 32 | 9 | 2 | 3 | 3 |

TABLE VI
TOP 5 NODES RANKING DOLPHIN

| *Rank* | *TR* | *BC* | *CNC* | *EC* | *PR* | *TC* |
|---|---|---|---|---|---|---|
| 1 | 15 | 37 | 37 | 15 | 52 | 46 |
| 2 | 46 | 2 | 41 | 38 | 18 | 15 |
| 3 | 34 | 41 | 38 | 46 | 58 | 38 |
| 4 | 14 | 38 | 21 | 34 | 34 | 34 |
| 5 | 58 | 8 | 15 | 51 | 30 | 14 |

TABLE VII
TOP 5 NODES RANKING BLOGS

| *Rank* | *TR* | *BC* | *CNC* | *EC* | *PR* | *TC* |
|---|---|---|---|---|---|---|
| 1 | 18 | 18 | 18 | 18 | 18 | 18 |
| 2 | 3 | 3 | 3 | 3 | 3 | 3 |
| 3 | 52 | 44 | 52 | 52 | 44 | 52 |
| 4 | 44 | 52 | 44 | 44 | 52 | 44 |
| 5 | 51 | 10 | 28 | 105 | 10 | 105 |

TABLE VIII
TOP 5 NODES RANKING USAIR97

| *Rank* | *TR* | *BC* | *CNC* | *EC* | *PR* | *TC* |
|---|---|---|---|---|---|---|
| 1 | 118 | 118 | 118 | 118 | 261 | 118 |
| 2 | 261 | 8 | 261 | 261 | 118 | 261 |
| 3 | 255 | 261 | 67 | 255 | 8 | 255 |
| 4 | 182 | 201 | 255 | 182 | 201 | 182 |
| 5 | 152 | 47 | 201 | 152 | 152 | 152 |

As shown in Table V top 5 nodes rank karate club, because in this paper we used a triangle neighborhood of nodes the ranking of the top 5 nodes of our proposed TC and TR (Triangle Centrality) rank the same node in the same order. But in the case of ranking the top rank node, the methods are divided into two groups, method TR, BC, CNC, and our proposed TC all rank node 1 as the top node. Meanwhile PR and EC rank node 34. As shown in Table VI the dolphins Network only a few of the measures rank the same node in the same ranking position. As shown in Table VII the Top 5 Nodes Ranking blogs network, here we can see that all the ranking methods rank nodes 18, 3 as the first and second top rank nodes. As shown in Table VIII the USAir97 Network the TR and our proposed TC top 5 rankings are the same but the top node in PR (node 262) and the others rank (node 119) as the top node.

We can now agree that different methods share different views of node power. In experiment 2 we evaluate the performance of our proposed method.

*Experiment 2: Density Impact of Removing Top 5 Rank Nodes.*

We compare the impact to a network graph density due to the removal of the top 5 rank nodes by several measures which include TR, BC, CNC, and our proposed TC.

$$dn = \frac{2N}{n(n-1)} \tag{5}$$

Where n is the number of nodes and N is the number of edges in G. The density is 0 for a graph without edges and 1 for a complete graph. The lower the density value, the higher the impact on the network. Figure 4 illustrates the result.

TABLE IX
NETWORK DENSITY VALUE

| | *Karate Club* | *Dolphins* | *Blogs* | *USAir97* |
|---|---|---|---|---|
| TR | 0.0468 | 0.0695 | 0.0492 | 0.0298 |
| BC | 0.0567 | 0.0783 | 0.0487 | 0.0323 |
| CNC | 0.0739 | 0.0739 | 0.0492 | 0.0306 |
| EC | 0.0468 | 0.0721 | 0.0480 | 0.0298 |
| PR | 0.0468 | 0.0714 | 0.0487 | 0.0316 |
| TC | 0.0468 | 0.0695 | 0.0480 | 0.0298 |

As shown in Table IX, the shaded cell represents the lowest value in a given column out of all the centrality methods used in this experiment our proposed method is the one with a consistently low density, which we can conclude that our proposed method (TC) outperform BC, CNC, EC, and PR based on our experiment. Figure 4 illustrates the plot generated from Table 9.

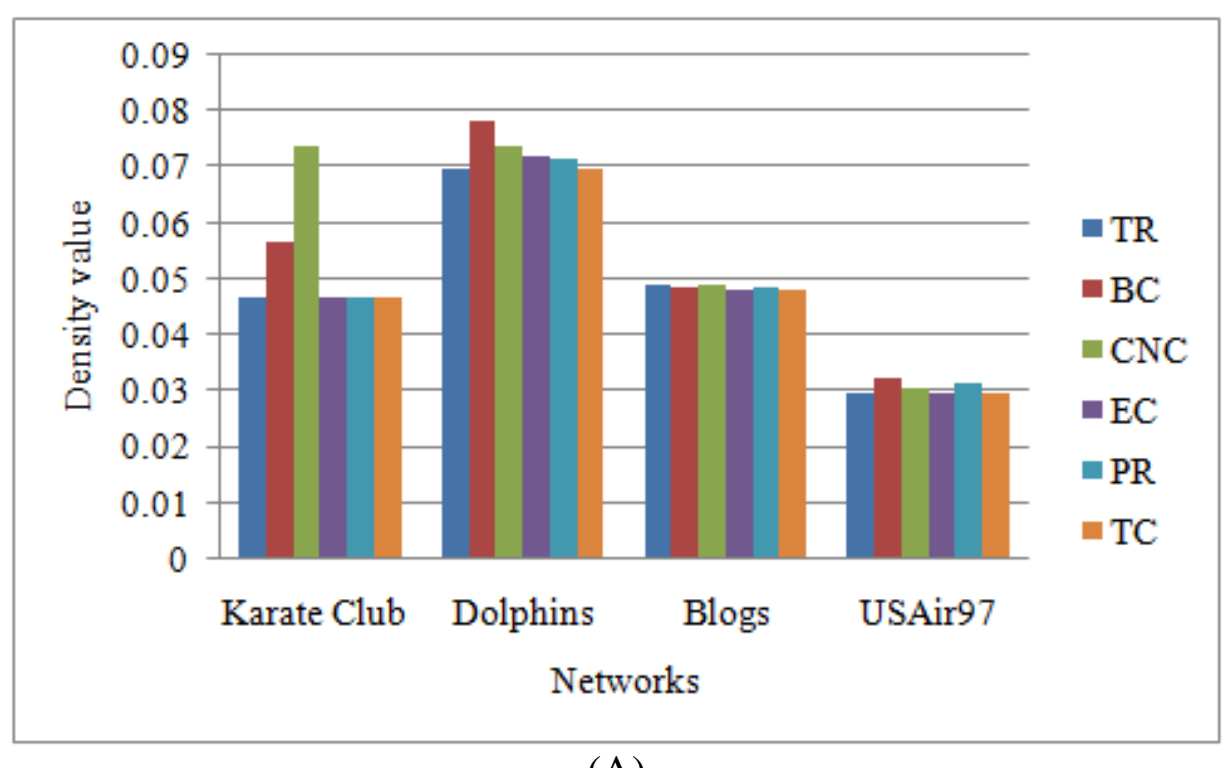

(A)

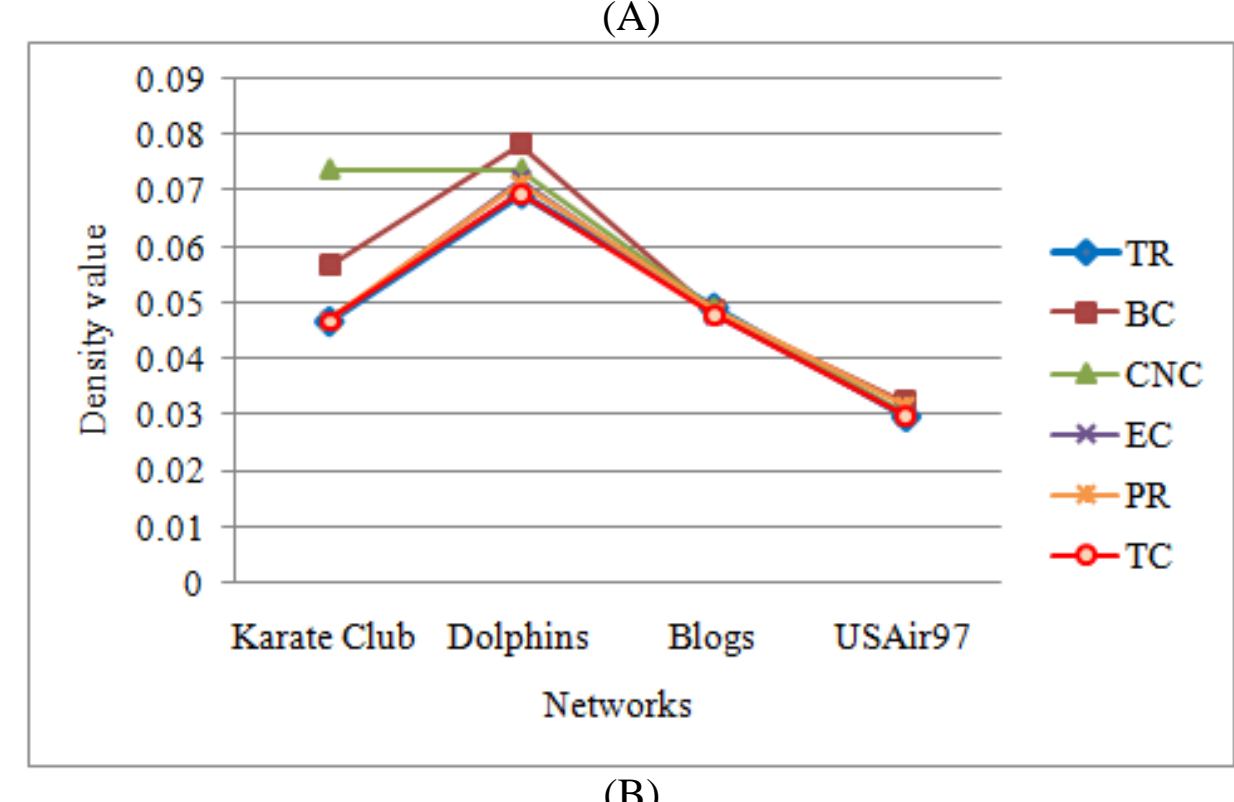

(B)

**Figure 4.** Network density value of experiment 2.

As shown in Figure 4, the x-axes represent networks which are the 1-Karate Club network, 2-Dolphins network, 3-Blogs Network, and 4- USAir97 network.

## V. CONCLUSION

In conclusion, this paper views complex network graphs as planar graphs and modifies Gruebler's equation to construct a new mathematical model (Tr-centrality) for the identification of node power in complex networks. We verified our Tr-centrality using four real network Datasets, after comparing the performance of our proposed equation with six well-known centrality methods, our proposed ranking equation yields a better result.

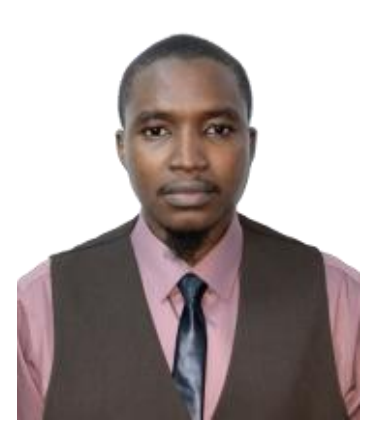

Auwal Tijjani Amshi (Graduate Member of GMCPN,) received BSc. degree in Computer science from Al-Hikmah University Ilorin, Nigeria in 2016, an MSc. degree in Software Engineering from Nanchang Hongkong University, China in 2020, and currently pursuing a PhD. degree. His research interests include Complex network analysis, Federated learning, and ITS.